\documentclass[%
 reprint,
 amsmath,amssymb,
 aps,
]{revtex4-2}

\usepackage{graphicx}
\usepackage{dcolumn}
\usepackage{bm}
\usepackage{xcolor}
\usepackage{bm}
\usepackage{hyperref}

\begin{document}

\preprint{APS/123-QED}

\title{Electromagnetic Scoot for Dyons Revisited}

\author{Rasim Yılmaz}%
\email{ryilmaz@metu.edu.tr}
\author{Onur Ayberk Çakmak}%
\email{acakmak@metu.edu.tr}
\affiliation{%
Department of Physics, Faculty of Arts and Sciences\\
Middle East Technical University, 06800, Ankara, Türkiye
}%

\date{\today}

\begin{abstract}
In the scattering of two electric charges, the particles acquire a shift in their net boost-like angular momentum, balanced by an opposite field contribution. This electromagnetic scoot effect appears at first order in post-Minkowskian expansion (1PM) order when conservation laws are evaluated on constant-time slices, but disappears at this order on hyperboloidal slices. Here, we extend this analysis to scattering involving both electric and magnetic charges and compare the results with the purely electric case in the context of multiparticle state representations.
\end{abstract}

\maketitle

\section{Introduction}
In the construction of an S-matrix, in- and out-states are described by multiparticle states, and the representations of these multiparticle states are generally taken as tensor products of one-particle states. Although this assumption seems consistent for the states at asymptotic times, in 1972 Zwanziger showed that in the scattering of electric and magnetic charges, an extra quantum number is necessary to fully describe the process \cite{zwanziger}. The observation Zwanziger made was that in this process, in addition to the orbital and spin angular momentum of each particle, there is also a residual angular momentum in the electromagnetic field of the in- and out-states. The generator associated to this angular momentum is added to the generators of the Lorentz transformations, and as a result, the scattering states do not transform as free particles. This modification has a simple group-theoretical description, and this description uses the idea of little group for pairs of particles. \\

In 2021, the result of Zwanziger on the multiparticle representations of the Poincare group was extended in a more systematic way \cite{Csaki1}. In this work, the authors extend the definition of asymptotic multiparticle states of the S-matrix, beyond the tensor products of one-particle states. This extension is made by generalizing the Wigner construction to the pairs of particles by introducing the so-called pairwise little groups. These pairwise little groups are defined as the subgroups of Lorentz transformations which leave a pair of momenta invariant. For a pair of particles, one can go to the center of momentum frame and the corresponding pairwise little group becomes $U(1)$. The authors call the charge related to this $U(1)$ transformation pairwise helicity. One can also see the topological origin of pairwise helicity in \cite{Mouland:2024zgk}.\\

Both \cite{zwanziger} and \cite{Csaki1} attributed this extra angular momentum and the corresponding quantum number to the scattering of dyons. However, in a recent work \cite{emscoot}, it has been shown that the extra contribution to the angular momentum is not limited to dyon-dyon scattering but can also be seen in the scattering of two electric charges. The authors of \cite{emscoot} considered the scattering of two electric charges and calculated the balance of the relativistic angular momentum to the leading order in perturbation. As a result, they showed that there is a shift of the net boost-like angular momentum or mass moment of the particles which is balanced by an opposite change of the boost-like angular momentum carried by the field, and they called this effect electromagnetic scoot. The electromagnetic scoot has been also calculated in a covariant fashion in \cite{Bhardwaj} and extended to gravitational scattering in \cite{selfforce}. This scoot effect implies that there should be an extra quantum number for multiparticle states related to boost transformation, and this boost-like quantum number has been calculated in \cite{boostquantumnumber}. We refer the reader to \cite{Lippstreu:2021avq} for further analysis and to \cite{Csaki:2020inw,Csaki:2021ozp,Csaki:2022ebw,Csaki:2022tvb} for some applications. \\

The same analysis on electromagnetic scoot was made with a different approach in \cite{Gralla:2024wzr}, where instead of using constant-$t$ slices, the conservation laws were formulated with hyperboloidal slices. It was found that when the electromagnetic mass moment is evaluated on a hyperboloidal slice $\tau =$ constant, the early-late time integrals vanish. Thus, the electromagnetic scoot effect at 1PM order is absent on this hyperboloidal boundary for the scattering of two electric charges. This raises a natural question: whether the same conclusion holds when magnetic charges are included, where residual field angular momentum already plays an essential role in the structure of asymptotic multiparticle states.\\

Motivated by this question, this work has two main parts. In the first part, we extend the analysis of \cite{emscoot} on the scattering of two electric charges to the scattering of two dyons. Zwanziger has already considered dyon-dyon scattering \cite{zwanziger}, and showed that there is a residual angular momentum in the electromagnetic field of the in- or out scattering states. Here, we obtain the same result following the analysis of \cite{emscoot}, and also calculate the change in the total mechanical angular momentum of the two particles to show the scoot effect. In Sect.\ref{sec2}, we consider a scattering encounter between two dyons in the approximation of small deflection and calculate their trajectories by using the Lorentz force law. In Sect.\ref{sec3}, we calculate conserved quantities at asymptotic times using these trajectories. As a result, we find the same extra contribution with \cite{zwanziger} to the angular momentum of electromagnetic field of in and out states, and show that this extra angular momentum carried by fields are balanced by an equal but opposite change in the total angular momentum of two dyons. Moreover, we calculate the extra boost-like angular momentum or mass moment carried by the fields at asymptotic times and the corresponding change in the total mass moment of the particles. \\

The second part focuses on the same analysis on the hyperboloidal slice boundary. Following \cite{Gralla:2024wzr}, we formulate the conservation laws with hyperboloidal slices for the scattering of an electric charge and a magnetic charge. In Sect.\ref{sec4a}, we discuss the field contribution to the mass moment and angular momentum. The mechanical contribution to the angular momentum is calculated in Sect.\ref{sec4b}. Our analysis shows that the scoot effect for the angular momentum is there for electric charge-magnetic charge scattering, even if one formulates the conservation laws on the hyperboloidal slice boundary.\\

\underline{\textbf{Conventions:}} Throughout this paper, we work in a $3+1$-dimensional Minkowski space embedded with the metric $\eta=\text{diag}(+1,-1,-1,-1)$. Our convention for the sign of the Levi-Civita symbol is  $\epsilon^{0123}=+1$.

\section{Small-angle Scattering of Relativistic Dyons} \label{sec2}
Following \cite{emscoot}, we consider the scattering of two dyons in the small deflection limit. We choose a frame in which one of the particles is initially at rest and denote this frame by a prime. The other dyon moves along the $z'$ direction and initial transverse separation is along the $x'$ direction. Then, the leading order trajectories are given by
\begin{equation}
    \boldsymbol{\text{r}}_1^{\prime}=\left(b, 0, v t^{\prime}\right), \quad \text{and} \quad \boldsymbol{\text{r}}_2^{\prime}=(0,0,0),
\label{leadingtrajectory}
\end{equation}
where $b$ is the impact parameter and $v$ is the relative velocity, which is chosen to be positive without loss of generality. Now, we present the Lorentz force acting on the particles due to the electric and magnetic fields created by one another. The most general form of the relativistic equation of motion for the $I$-th particle, due to fields generated by the $J$-th particle, is given by:
\begin{equation}
    \frac{d}{d t}(\gamma_I m_I \mathbf{v}_I)=e_I[\mathbf{E}_J+\mathbf{v}_I \times \mathbf{B}_J]+g_I[\mathbf{B}_J-\mathbf{v}_I \times \mathbf{E}_J],
\end{equation}
where $e_I$ denotes the electric charge, $g_I$ denotes the magnetic charge and $\gamma_I=(1-\mathbf{v}_I\cdot\mathbf{v}_I)^{-1/2}$ is the Lorentz factor. For a two particle system, the corresponding equations written in a generic frame read
\begin{equation}
    \begin{aligned}
        &\frac{d}{d t}(\gamma_1 m_1 \mathbf{v}_1)=e_1[\mathbf{E}_2+\mathbf{v}_1 \times \mathbf{B}_2]+g_1[\mathbf{B}_2-\mathbf{v}_1 \times \mathbf{E}_2], \\
        &\frac{d}{d t}(\gamma_2 m_2 \mathbf{v}_2)=e_2[\mathbf{E}_1+\mathbf{v}_2 \times \mathbf{B}_1]+g_2[\mathbf{B}_1-\mathbf{v}_2 \times \mathbf{E}_1].
    \end{aligned}
\end{equation}
Using the leading order trajectories in the primed frame \eqref{leadingtrajectory}, the equations take the form
\begin{equation}
    \begin{aligned}
        m_1 \gamma\left[\ddot{\mathbf{r}}'_1+{\gamma^2} v \hat{\mathbf{z}}'\left(v \hat{\mathbf{z}}' \cdot \ddot{\mathbf{r}}'_1\right)\right]&=e_1(\mathbf{E}'_2+v \hat{\mathbf{z}}' \times \mathbf{B}'_2)\\
         &\quad +g_1(\mathbf{B}'_2-v \hat{\mathbf{z}}' \times \mathbf{E}'_2), \\
        m_2\ddot{\mathbf{r}}'_2 &=e_2\mathbf{E}'_1+g_2\mathbf{B}'_1.
    \end{aligned}
 \label{lorentzforcelaws}   
\end{equation}
The electric and magnetic fields generated by the second particle at an arbitrary position are given by
\begin{equation}
    \begin{aligned}
    \mathbf{E}_2^{\prime} &=e_2 \frac{x^{\prime} \hat{\mathbf{x}}'+y^{\prime} \hat{\mathbf{y}}'+z^{\prime} \hat{\mathbf{z}}'}{\left[x^{\prime 2}+y^{\prime 2}+z^{\prime 2}\right]^{3 / 2}}, \\ \mathbf{B}_2^{\prime} &=g_2 \frac{x^{\prime} \hat{\mathbf{x}}'+y^{\prime} \hat{\mathbf{y}}'+z^{\prime} \hat{\mathbf{z}}'}{\left[x^{\prime 2}+y^{\prime 2}+z^{\prime 2}\right]^{3 / 2}},
    \end{aligned}
\end{equation}
and for the first particle they are given by
\begin{equation}
\begin{aligned}
    & \mathbf{E}_1^{\prime}=\gamma(\mathbf{E}_1-\mathbf{v}_1 \times \mathbf{B}_1)-\frac{\gamma^2}{\gamma+1} (\mathbf{v}_1 \cdot \mathbf{E}_1)\mathbf{v}_1, \\
    & \mathbf{B}_1^{\prime}=\gamma(\mathbf{B}_1+\mathbf{v}_1 \times \mathbf{E}_1)-\frac{\gamma^2}{\gamma+1}(\mathbf{v}_1 \cdot \mathbf{B}_1)\mathbf{v}_1,
\end{aligned}
\label{lorentztransformsoffields}
\end{equation}
where $\mathbf{E}_1$ and $\mathbf{B}_1$ are electric and magnetic fields of particle 1 in its rest frame and are given by
\begin{equation}
\begin{aligned}
        &\mathbf{E}_1=e_1 \frac{x\hat{\mathbf{x}}+y \hat{\mathbf{y}}+z \hat{\mathbf{z}}}{\left[x^2+y^{ 2}+z^2\right]^{3 / 2}} \\
        &\mathbf{B}_1=g_1 \frac{x \hat{\mathbf{x}}+y \hat{\mathbf{y}}+z \hat{\mathbf{z}}}{\left[x^2+y^{ 2}+z^2\right]^{3 / 2}}.
\end{aligned}
\label{fieldsinframe1}
\end{equation}
Inserting (\ref{fieldsinframe1}) into (\ref{lorentztransformsoffields}) and using the Poincar\'{e} transformations for the coordinates,
\begin{equation}
    x = x'-b, \quad y=y', \quad z=\gamma (z'-vt'), \quad t= \gamma (t'-vz'),
\end{equation}
the electric-magnetic fields $\mathbf{E}'_1$ and $\mathbf{B}'_1$ in the primed frame become 
\begin{equation}
\begin{aligned}
    &\mathbf{E}'_1 = \frac{\gamma[e_1\bar{x}+vg_1y']\hat{\mathbf{x}}'+\gamma[e_1y'-vg_1\bar{x}]\hat{\mathbf{y}}'+e_1 \gamma k\hat{\mathbf{z}}'}{\left[\bar{x}^2+y'^{ 2}+\gamma^2k^2\right]^{3 / 2}}, \\
    &\mathbf{B}'_1 = \frac{\gamma[g_1\bar{x}-ve_1y']\hat{\mathbf{x}}'+\gamma[g_1y'+ve_1\bar{x}]\hat{\mathbf{y}}'+g_1 \gamma k\hat{\mathbf{z}}'}{\left[\bar{x}^2+y'^{ 2}+\gamma^2k^2\right]^{3 / 2}},
\end{aligned}\label{E1_B1}
\end{equation}
where $\bar{x}=(x'-b)$ and $k=(z'-vt')$. Evaluating \eqref{E1_B1} at $\mathbf{r}_2^{\prime}=(0,0,0)$ one gets
\begin{equation}
\begin{aligned}
           &\mathbf{E}'_1=\frac {-e_1 \gamma b\hat{\mathbf{x}}'+g_1 v \gamma b \hat{\mathbf{y}}'-e_1 \gamma vt' \hat{\mathbf{z}}' }{(b^2+{\gamma}^2v^2 t'^2)^{3 / 2}}, \\
           &\mathbf{B}'_1= \frac {-g_1 \gamma b\hat{\mathbf{x}}'-e_1 v \gamma b \hat{\mathbf{y}}'-g_1 \gamma vt' \hat{\mathbf{z}}'  }{(b^2+{\gamma}^2v^2 t'^2)^{3 / 2}}.
\end{aligned}
\label{em1atr2}
\end{equation}
Similarly, the electric and magnetic fields $\mathbf{E}'_2$ and $\mathbf{B}'_2$ in the primed frame, evaluated at $\mathbf{r}_1^{\prime}=\left(b, 0, v t^{\prime}\right)$, take the form
\begin{equation}
    \mathbf{E}_2^{\prime}=e_2 \frac{b \hat{\mathbf{x}}^{\prime}+vt' \hat{\mathbf{z}}^{\prime}}{(b^2+v^2t'^2)^{3/2}} \quad \text{and} \quad \mathbf{B}_2^{\prime}=g_2 \frac{b \hat{\mathbf{x}}^{\prime}+ vt' \hat{\mathbf{z}}^{\prime}}{(b^2+v^2t'^2)^{3/2}}.
\label{em2atr1}
\end{equation}
Inserting (\ref{em1atr2}) and (\ref{em2atr1}) into (\ref{lorentzforcelaws}), one obtains the accelerations
\begin{equation}
    \begin{aligned}
        &a'_{x1} =\frac{(q_1q_2)b}{m_1 \gamma (b^2+v^2t'^2)^{3/2}}, \quad a'_{y1} = \frac{(e_1 g_2-e_2g_1)vb}{m_1 \gamma (b^2+v^2t'^2)^{3/2}}, \\
        &a'_{z1}=\frac{(q_1q_2)vt'}{m_1 \gamma^3 (b^2+v^2t'^2)^{3/2}},\\
        &a'_{x2}= \frac{-(q_1q_2)\gamma b}{m_2(b^2+{\gamma}^2v^2 t'^2)^{3 / 2}},  \quad a'_{y2} = \frac{-(e_1 g_2-e_2g_1)\gamma vb}{m_2(b^2+{\gamma}^2v^2 t'^2)^{3 / 2}},\\
        &a'_{z2} = \frac{-(q_1q_2)\gamma vt'}{m_2(b^2+{\gamma}^2v^2 t'^2)^{3 / 2}},
    \end{aligned}
\label{eoms}
\end{equation}
where we defined $q_1q_2 := e_1e_2+g_1g_2$. Integrating both sides of the equations in (\ref{eoms}), we obtain the trajectories for the first particle as
\begin{equation}
\begin{aligned}
    &x'_1=b+\frac{(q_1q_2)}{m_1 \gamma b v^2}(\sqrt{b^2+v^2t'^2}+vt'), \\ 
    &y'_1= \frac{(e_1g_2-g_1e_2)}{m_1 \gamma b v}(\sqrt{b^2+v^2t'^2}+vt'), \\
    &z'_1 =v t^{\prime}-\frac{q_1q_2}{m_1 \gamma^3 v^2} \operatorname{arctanh} \left(\frac{v t^{\prime}}{\sqrt{v^2 t^{\prime 2}+b^2}}\right),
\end{aligned}
\label{trajectory1}
\end{equation}
and for the second particle as
\begin{equation}
   \begin{aligned}
       &x_2^{\prime}=-\frac{(q_1q_2)}{b m_2 v^2}\left(v t^{\prime}+\sqrt{v^2 t^{\prime 2}+b^2 \gamma^{-2}}\right), \\
       &y_2^{\prime}= -\frac{(e_1 g_2-g_1e_2)}{b m_2 v}\left(v t^{\prime}+\sqrt{v^2 t^{\prime 2}+b^2 \gamma^{-2}}\right),\\
       &z_2^{\prime}=\frac{(q_1q_2)}{m_2 \gamma^2 v^2} \operatorname{arctanh} \left( \frac{v t^{\prime}}{\sqrt{v^2 t^{\prime 2}+b^2 \gamma^{-2}}} \right).
   \end{aligned}
\label{trajectory2}
\end{equation}
When integrating (\ref{eoms}) to obtain (\ref{trajectory1}) and (\ref{trajectory2}), we used the initial conditions
\begin{equation}
    t' \rightarrow -\infty: \quad x'_1 \rightarrow b, \quad y'_1 \rightarrow 0, \quad x'_2 \rightarrow 0, \quad y'_2 \rightarrow 0
\end{equation}
Following \cite{emscoot}, we transition to the center of momentum (COM) frame, before we begin to evaluate the conserved quantities. Introducing
\[
    E_0:=\sqrt{m_1^2+m_2^2+2\gamma m_1m_2},
\]
the corresponding transformations to the COM frame are given by
\begin{equation}
\begin{aligned}
    &t=\frac{m_2+\gamma m_1}{E_0} t^{\prime}-\frac{\gamma m_1 v}{E_0} z^{\prime}, \\
    &z=\frac{m_2+\gamma m_1}{E_0} z^{\prime}-\frac{\gamma m_1 v}{E_0} t^{\prime}, \\
    &x=x^{\prime}-b \frac{m_1\left(\gamma m_2+m_1\right)}{E_0^2}, \\
    &y=y'.
\end{aligned}
\end{equation}
Using these transformations, the trajectories (\ref{trajectory1}) for the first particle become
\begin{equation}\label{eq:pos1}
\begin{aligned}
    &x_1=b_1+\frac{(q_1q_2)}{b m_1 \gamma v^2}\left(v t_1+\sqrt{b^2+v^2 t_1^2}\right), \\
    &y_1= \frac{(e_1g_2-g_1e_2)}{b m_1 \gamma v}(vt_1+\sqrt{b^2+v^2t_1^2}),\\
    &z_1=v_1\left(t-\frac{(q_1q_2) E_0}{m_1 m_2 \gamma^3 v^3} \operatorname{arctanh} \frac{v t_1}{\sqrt{b^2+v^2 t_1^2}}\right),
\end{aligned}
\end{equation}
and the trajectories (\ref{trajectory2}) for the second particle become
\begin{equation}\label{eq:pos2}
    \begin{aligned}
        &x_2=b_2-\frac{(q_1q_2)}{b m_2 \gamma v^2}\left(v t_2+\sqrt{b^2+v^2 t_2^2}\right), \\
        &y_2=-\frac{(e_1 g_2-g_1e_2)}{b m_2 v}\left(v t_2+\sqrt{v^2 t_2^{ 2}+b^2 \gamma^{-2}}\right), \\
        &z_2=v_2\left(t-\frac{(q_1q_2) E_0}{m_1 m_2 \gamma^3 v^3} \operatorname{arctanh} \frac{v t_2}{\sqrt{b^2+v^2 t_2^2}}\right),
    \end{aligned}
\end{equation}
where the parameters are given as
\begin{equation}
    \begin{aligned}
        t_1 & =\frac{\gamma E_0}{m_1+\gamma m_2} t, & t_2 & =\frac{\gamma E_0}{m_2+\gamma m_1} t \\
        b_1 & =\frac{m_2\left(m_2+\gamma m_1\right)}{E_0^2} b, & b_2 & =-\frac{m_1\left(m_1+\gamma m_2\right)}{E_0^2} b \\
        v_1 & =\frac{\gamma m_2}{m_1+\gamma m_2} v, & v_2 & =-\frac{\gamma m_1}{m_2+\gamma m_1} v .
    \end{aligned}
\end{equation}\\

In this section, we have extended the results of electron-electron scoot studied in \cite{emscoot} to a generic dyon-dyon case. The trajectories of the particles in the COM frame were calculated and presented in \eqref{eq:pos1} and \eqref{eq:pos2}. In the next section, we compute the conserved quantities due to particle and field contributions related to the aforementioned scoot.  

\section{Conserved Quantities}\label{sec3}
\subsection{Particle Contribution}
In this section we calculate the conserved energy, momentum, angular momentum, and mass moment at asymptotic times. These quantities for the particles are given as
\begin{equation}
    \begin{aligned}
        E_I & =\gamma_I m_I, \\
        \mathbf{p}_I & =\gamma_I m_I {\dot{\mathbf{r}}_I}, \\
        \mathbf{L}_I & =\mathbf{r}_I \times \mathbf{p}_I, \\
\mathbf{N}_I & =\gamma_I m_I \mathbf{r}_I-\mathbf{p}_I t,
\end{aligned}
\label{conservedquant}
\end{equation}
where $I=1,2$ is the particle index. The position vectors of the particles at asymptotic times can be found by evaluating \eqref{eq:pos1} and \eqref{eq:pos2} in the limit $t\to \pm\infty$\footnote{In contrast to \cite{emscoot}, we find a $\mathcal{O}(t^{-1})$ dependence for the higher-order terms in $x_I$ and $y_I$.}
\begin{align} 
    x_I &= b_I + \Theta(t) \dfrac{2(q_1q_2)}{\mu b\gamma v^2} v_It + \mathcal{O}(t^{-1}) \label{eq:xI} \\
    y_I &= -\Theta(t)\dfrac{2(e_1g_2-e_2g_1)}{\mu b\gamma v}v_It  + \mathcal{O}(t^{-1}) \label{eq:yI} \\
    z_I &= v_I
    \left( 
        t-\text{sgn}(t) \dfrac{q_1q_2}{\mu\gamma^3v^3} \log \dfrac{2v|t_I|}{b}
    \right) + \mathcal{O}(t^{-2}), \label{eq:zI}
\end{align}
Plugging \eqref{eq:xI}-\eqref{eq:zI} into (\ref{conservedquant}) yields 
\begin{align}
    E_1 =& \dfrac{m_1+\gamma m_2}{E_0}
    \left( 
        m_1-\dfrac{m_2}{E_0}\dfrac{q_1q_2}{\gamma v|t|}
    \right) + \mathcal{O}(t^{-2}), \\
    \mathbf{p}_1 =& +\Theta(t) \dfrac{2(q_1q_2)}{bv}\hat{\mathbf{x}} - \Theta(t)\dfrac{2(e_1g_2-e_2g_1)}{b}\mathbf{\hat{y}} \nonumber \\
        & + 
    \left( 
        \mu\gamma v - \dfrac{q_1q_2}{\gamma^2v^2}\dfrac{(m_1+\gamma m_2)^2}{E_0^2|t|}
    \right)\hat{\mathbf{z}} + \mathcal{O}(t^{-2}), \\
    \mathbf{L}_1 =&  - \mu b\gamma v \dfrac{m_2(m_2+\gamma m_1)}{E_0^2}\hat{\mathbf{y}} \nonumber \\
    &- \Theta(t) 2(e_1g_2-e_2g_1) \dfrac{m_2(m_2+\gamma m_1)}{E_0^2}\hat{\mathbf{z}} + \mathcal{O}(t^{-1}), \\
    \mathbf{N}_1 =& \gamma_1m_1b_1\hat{\mathbf{x}} - \text{sgn}(t) \dfrac{q_1q_2}{\gamma^2v^2}
    \left( 
        \log\dfrac{2\gamma vE_0|t|}{(m_1+\gamma m_2)b} - 1 
    \right)\hat{\mathbf{z}} \nonumber \\
    &+ \mathcal{O}(t^{-1}),
\end{align}
and 
\begin{align}
    E_2 =& \dfrac{m_2+\gamma m_1}{E_0}
    \left( 
        m_2-\dfrac{m_1}{E_0}\dfrac{q_1q_2}{\gamma v|t|}
    \right) + \mathcal{O}(t^{-2}), \\
    \mathbf{p}_2 =& -\Theta(t) \dfrac{2(q_1q_2)}{bv}\hat{\mathbf{x}} + \Theta(t)\dfrac{2(e_1g_2-e_2g_1)}{b}\hat{\mathbf{y}} \nonumber \\
        &\ - 
    \left( 
        \mu\gamma v - \dfrac{q_1q_2}{\gamma^2v^2}\dfrac{(m_2+\gamma m_1)^2}{E_0^2|t|}
    \right)\hat{\mathbf{z}} + \mathcal{O}(t^{-2}), \\
    \mathbf{L}_2 =&  - \mu b\gamma v \dfrac{m_1(m_1+\gamma m_2)}{E_0^2}\hat{\mathbf{y}} \nonumber \\
    &- \Theta(t) 2(e_1g_2-e_2g_1) \dfrac{m_1(m_1+\gamma m_2)}{E_0^2}\hat{\mathbf{z}} + \mathcal{O}(t^{-1}), \\
    \mathbf{N}_2 =& \gamma_2m_2b_2\hat{\mathbf{x}} - \text{sgn}(t) \dfrac{q_1q_2}{\gamma^2v^2}
    \left( 
        \log\dfrac{2\gamma vE_0|t|}{(m_2+\gamma m_1)b} - 1
    \right)\hat{\mathbf{z}} \nonumber \\
    &+ \mathcal{O}(t^{-1}).
\end{align}
Summing them, the total contributions from the two particles become
\begin{align}
    E_1+E_2  =&E_0-\frac{(q_1q_2)}{v|t|} \frac{m_1^2+m_2^2+2 m_1 m_2 / \gamma}{E_0^2} + \mathcal{O}(t^{-2}), \\
    \mathbf{p}_1+\mathbf{p}_2  =&-\frac{(q_1q_2)}{|t|} \frac{m_2^2-m_1^2}{E_0^2} \hat{\mathbf{z}}+O\left(t^{-2}\right), \\
    \mathbf{L}_1+\mathbf{L}_2  =&-\mu b \gamma v \hat{\mathbf{y}} + 2(e_1g_2-e_2g_1)\Theta(t)\hat{\mathbf{z}}+O\left(t^{-1}\right), \\
    \mathbf{N}_1+\mathbf{N}_2  =& - \text{sgn}(t) \frac{(q_1q_2)}{\gamma^2 v^2} \log \frac{m_2+\gamma m_1}{m_1+\gamma m_2} \hat{\mathbf{z}}+O\left(t^{-1}\right).
\end{align}
Considering these quantities at initial and final times, the total changes in particle contributions read
\[
    \Delta E_{\mathrm{mech}}=0, \quad \Delta \mathbf{p}_{\mathrm{mech}}=0,
\]
and
\begin{equation}
    \begin{aligned}
        \Delta \mathbf{N}_{\mathrm{mech}}=&-\frac{2 (q_1q_2)}{\gamma^2 v^2} \log \frac{m_2+\gamma m_1}{m_1+\gamma m_2} \hat{\mathbf{z}}, \\
     \Delta \mathbf{L}_{\mathrm{mech}}=&\; 2(e_1g_2-e_2g_1)\hat{\mathbf{z}}.
    \end{aligned}
 \label{mechchange}   
\end{equation}
The first difference from the findings of \cite{emscoot} is in $\Delta \mathbf{L}_\text{mech}$, which vanishes in the case of electron-electron scattering ($g_1=g_2=0)$. Also note that in our case $\Delta \mathbf{N}_{\mathrm{mech}}$ takes contribution from both electric and magnetic charges, remembering our definition $q_1q_2 := e_1e_2+g_1g_2$.
\subsection{Field Contribution}
As discussed in \cite{emscoot}, the leading electromagnetic field is determined by the leading straight-line motion of the charges
\begin{equation}
    \begin{aligned}
\mathbf{E}_I & =\frac{ \gamma_I}{R_I^3}\bigg(e_I\left(x-b_I\right)+v_Ig_Iy, \, e_Iy-g_Iv_I(x-b_I), \,e_I k_I \bigg), \\
\mathbf{B}_I & =\frac{ \gamma_I}{R_I^3}\bigg(g_I(x-b_I)-e_Iv_I y, \, g_Iy+e_I v_I (x-b_I), \,g_I k_I\bigg),
\end{aligned}
\end{equation}
where $\gamma_I=\left(1-v_I^2\right)^{-1 / 2}$ and
\begin{equation}
    R_I=\sqrt{\left(x-b_I\right)^2+y^2+\gamma_I^2\left(z-v_I t\right)^2}.
\end{equation}
The field contributions to the conserved quantities are given by
\begin{equation}
\begin{aligned}
    E_F & =\frac{1}{8 \pi} \int d^3 x\left(\mathbf{E}^2+\mathbf{B}^2\right)  ,\\
    \mathbf{p_F} & =\frac{1}{4 \pi} \int d^3 x(\mathbf{E} \times \mathbf{B}) ,\\
    \mathbf{L_F} & =\frac{1}{4 \pi} \int d^3 x \left( \mathbf{x} \times(\mathbf{E} \times \mathbf{B}) \right),\\
    \mathbf{N_F} & = \frac{1}{8 \pi} \int d^3 x\left(\mathbf{E}^2+\mathbf{B}^2\right) \mathbf{x} -\mathbf{p}_F t.
\end{aligned}
\label{fieldconserved}
\end{equation}
Eq (\ref{fieldconserved}) contains both self-field contributions and cross-term contributions to the conserved quantities. Following \cite{emscoot}, we only take the cross term contributions, and hence (\ref{fieldconserved}) becomes
\begin{equation}\label{eq:cross_terms}
\begin{aligned}
    E_{F \times} & =\frac{1}{2} \int d^3 x \ \mathcal{E}_{\times} , \\
    \mathbf{p_{F \times}} & =  \int d^3 x \ \mathbf{S}_{\times}, \\
    \mathbf{L_{F \times}} & =  \int d^3 x \ (\mathbf{x} \times \mathbf{S}_{\times}) ,\\
    \mathbf{N_{F \times}} & =-\mathbf{p}_{F \times} t +\frac{1}{2} \int d^3 x \ (\mathcal{E}_{\times} \mathbf{x}),
\end{aligned}
\end{equation}
where we defined
\begin{equation}\label{eq:E_S}
\begin{aligned}
    & \mathcal{E}_{\times}=\frac{1}{4 \pi}\left(\mathbf{E}_1 \cdot \mathbf{E}_2+\mathbf{B}_1 \cdot \mathbf{B}_2\right), \\
    & \mathbf{S}_{\times}=\frac{1}{4 \pi}\left(\mathbf{E}_1 \times \mathbf{B}_2+\mathbf{E}_2 \times \mathbf{B}_1\right) .
\end{aligned}
\end{equation}
Setting $b=0$ and changing to cylindrical coordinates, the electric and magnetic fields take the form
\begin{equation}
    \begin{aligned}
        & \tilde{\mathbf{E}}_I=\frac{ \gamma_I}{\tilde{R}_I^3}
        \left[
            \ e_I\rho \hat{\bm{\rho}}-  g_I v_I \rho \hat{\bm{\phi}}+  e_I k_I \hat{\mathbf{z}}\right] \\
            & \tilde{\mathbf{B}}_I=\frac{ \gamma_I}{\tilde{R}_I^3} \left[g_I \rho \hat{\bm{\rho}} +e_I v_I \rho \hat{\bm{\phi}}  +g_I k_I \hat {\mathbf{z}}  \right ], 
\end{aligned}
\end{equation}
where $k_I = z- v_It$ and,
\begin{equation}
    \tilde{R}_I = \sqrt{\rho^2+\gamma_I^2k_I^2}.
\end{equation}
Quantities in \eqref{eq:E_S} expressed in cylindriEcal coordinates read 
\begin{equation}\label{eq:E_S2}
\begin{aligned}
    \mathcal{E}_\times &= \dfrac{(e_1e_2+g_1g_2)}{4\pi} \dfrac{\gamma_1\gamma_2[k_1k_2+(1+v_1v_2)\rho^2]}{\tilde{R}_1^{3}\tilde{R}^{3}_2}, \\
    \mathbf{S}_\times &= (\dots \sin\phi + \dots \cos\phi)\hat{\mathbf{x}} +(\dots \sin\phi + \dots \cos\phi)\hat{\mathbf{y}} \\
        &\quad + \dfrac{e_1e_2+g_1g_2}{4\pi}\dfrac{\gamma_1\gamma_2(v_1+v_2)\rho^2}{\tilde{R}^3_1\tilde{R}^3_2} \hat{\mathbf{z}},
\end{aligned}
\end{equation}
where ``$\dots$" represents coefficients independent of the variable $\phi$. 
Substituting \eqref{eq:E_S2} into  \eqref{eq:cross_terms}, the field contributions at infinite past and infinite future (up to linear order contribution) are
\begin{equation}\label{eq:cons}
\begin{aligned}
    E_{F\times} &= \dfrac{e_1e_2+g_1g_2}{E_0^2|vt|}
    \left(
        m_1^2+m_2^2+2\dfrac{m_1m_2}{\gamma}
    \right), \\
    \mathbf{p}_{F\times} &=  \dfrac{e_1e_2+g_1g_2}{|t|}\dfrac{m^2_2-m_1^2}{E_0^2}\hat{\mathbf{z}}, \\
    \mathbf{L}_{F\times} &= -\text{sgn}(t)(e_1g_2-e_2g_1)\hat{\mathbf{z}} , \\
    \mathbf{N}_{F\times}&= \pm \dfrac{e_1e_2+g_1g_2}{\gamma^2v^2}\log \dfrac{m_2+\gamma m_1}{m_1+\gamma m_2}\hat{\mathbf{z}}.
\end{aligned}
\end{equation}
Given that the field contribution of $\mathbf{L}_{F\times}$ vanishes in \cite{emscoot}, we explicitly present our calculations here 
\begin{equation}\label{eq:LF}
\begin{aligned}
    \mathbf{L}_{F\times} &= \int d^3x (\mathbf{x}\times \mathbf{S_\times})  \\
        &= \int\limits_{-\infty}^\infty dz \int\limits_0^\infty \rho d\rho \int\limits_0^{2\pi} d\phi \\
            &\qquad \times 
            \Big[
                \hat{\mathbf{x}}(\dots \sin\phi + \dots\cos\phi) \\
                &\qquad\quad + \hat{\mathbf{y}}(\dots\sin\phi + \dots\cos\phi) \\
                &\qquad\quad + \hat{\mathbf{z}}\dfrac{(e_1g_2-e_2g_1)}{4\pi}\dfrac{\gamma_1\gamma_2(k_1-k_2)\rho^2}{\tilde{R}_1^3\tilde{R}_2^3}
            \Big] \\
        &= \dfrac{e_1g_2-e_2g_1}{2}\hat{\mathbf{z}}\int\limits_{-\infty}^\infty  \dfrac{\gamma_1\gamma_2(k_1-k_2)}{(s_1\gamma_1k_1+s_2\gamma_2k_2)^2}dz,
\end{aligned}
\end{equation}
where $s_I\equiv \text{sgn}({k}_I)$. \\

To evaluate the last line of \eqref{eq:LF}, it is useful to make the substitution $z\to z|t|$ and evaluate the limit $t\to \pm \infty$:
\begin{equation}\label{eq:zt}
    \mathbf{L}_{F\times} = \mp\hat{\mathbf{z}}\int\limits_{\infty}^\infty dz \ \dfrac{(e_1g_2-e_2g_1)(v_1-v_2)}{2[\tilde{s}_1\gamma_1(z\mp v_1)+\tilde{s}_2\gamma_2(z\mp v_2)]^2}.
\end{equation}
Here, the upper (lower) sign corresponds to the $t\to+\infty\;(-\infty)$ limit and we have defined $\tilde{s}_I\equiv \text{sgn}(z\mp v_I)$. The integral \eqref{eq:zt} has poles at $z=\pm v_1$ and $z=\pm v_2$ and hence should be divided into three regions. The final result turns out to be 
\begin{equation}
    \mathbf{L_{F\times}} = \mp (e_1g_2-e_2g_1)\hat{\mathbf{z}},
\end{equation}
which is identical to that in \eqref{eq:cons}. This implies that the change in the angular momentum of the electromagnetic field is given by
\begin{equation}
    \Delta \mathbf{L_{F\times}} = -2(e_1g_2-e_2g_1)\hat{\mathbf{z}},
\end{equation}
which balances the mechanical change in the angular momentum given in (\ref{mechchange}). So, we have extended the results of \cite{emscoot} on the electromagnetic scoot effect to the scattering of two dyons.

\section{1PM Scoot on the hyperboloidal slice boundary}
\subsection{Field Contribution}\label{sec4a}
Until now, we have followed \cite{emscoot} and extended the results found in that work to dyon scattering. In another interesting work \cite{Gralla:2024wzr}, the authors examine the conserved quantities in an electromagnetic scattering process using a different approach, namely, they transition from constant $t$-slices and formulate conservation laws with hyperboloidal slices. They show that one does not see the scoot effect at 1PM order when the conserved quantities are evaluated on the hyperboloidal slice boundary. We refer the reader to \cite{Gralla:2024wzr} for a detailed discussion on the hyperboloidal slice boundary. In this section, we extend this analysis to the dyon scattering and show that there is still a scoot effect for the angular momentum. \\

Following \cite{Gralla:2024wzr}, the total mass moment on a constant-$\tau$ surface is given as
\begin{equation}
    N_{(\tau)}^i=\tau^2 \int d \rho d \theta d \phi \sinh ^2 \rho \sin \theta\left(T^{0 \nu} x^i-T^{i \nu} t \right)x_\nu.
\label{massmoment}
\end{equation}
The corresponding expression for the angular momentum is
\begin{equation}
    M_{(\tau)}^{ij}=\tau^2 \int d \rho d \theta d \phi \sinh ^2 \rho \sin \theta\left(T^{i \nu}  x^j-T^{j \nu} x^i \right)x_\nu,
\label{angmom}
\end{equation}
where 
\begin{equation}
    t=\tau \text{cosh} \rho, \quad r=\tau \text{sinh}\rho,
    \label{puzzle_coords}
\end{equation}
with $x^\nu=(t,x,y,z)$. To calculate these quantities, we need to construct the electromagnetic energy-momentum tensor using the definition
\begin{equation}
    T^{\mu \nu}=F^{\mu \alpha} F^\nu{}_\alpha-\frac{1}{4} \eta^{\mu \nu} F_{\alpha \beta} F^{\alpha \beta}.
\label{emstrestensor}
\end{equation}
The components take the form
\begin{equation}
    \begin{aligned}
        T^{00}&= T^{tt}= \frac{1}{2}\left(\mathbf{E}^2+\mathbf{B}^2 \right), \\
        T^{i0}&= T^{it}= (\mathbf{E} \times \mathbf{B})^i,\\
        T^{ij}&=- E^i E^j- B^i B^j+\frac{1}{2}\left( \mathbf{E}^2+ \mathbf{B}^2\right) \delta^{i j}.
    \end{aligned}
\label{emtensorcomps}
\end{equation}
The electric and magnetic fields of point particles moving along some specified world-lines $\mathbf{r}_I(t)$ with velocity $\mathbf{v}_I$ and acceleration $\mathbf{a}_I$ are given by 
\begin{equation}
    \begin{aligned}
        \mathbf{E}_{C, I}=&\left.\frac{q_I}{R_I^2} \frac{\hat{\mathbf{R}}_I-\mathbf{v}_I}{\left(1-v_I^2\right)^{-1}\left(1-\hat{\mathbf{R}}_I \cdot \mathbf{v}_I\right)^3}\right|_{\mathrm{ret}}, \\
       \mathbf{E}_{R, I}=&\left.\frac{q_I}{R_I} \frac{\hat{\mathbf{R}}_I \times\left((\hat{\mathbf{R}}_I-\mathbf{v}_I) \times \mathbf{a}_I\right)}{\left(1-\hat{\mathbf{R}}_I \cdot \mathbf{v}_I\right)^3}\right|_{\mathrm{ret}}, \\
       \mathbf{B}_{ I}=&\left.\hat{\mathbf{R}}_I\right|_{\mathrm{ret}} \times \mathbf{E}_I,
    \end{aligned}
\label{ebfields}
\end{equation}
where 
\begin{equation}
    \mathbf{R}_I(t)=r-r_I(t), \quad \hat{\mathbf{R}}_I=\mathbf{R}_I / R_I,
\label{Rvector}
\end{equation}
and subscript “ret” indicates that the fields are evaluated at the retarded time $t_r$ that satisfies $R_I\left(t_r\right)=t-t_r $. Assuming that the particles are widely separated at late times, their velocities, positions, and accelerations take the form
\begin{equation}
    \begin{aligned}
        \mathbf{v}_I=&\mathbf{V}_I+O(1 / t), \\
        \mathbf{r}_I=&\mathbf{V}_It+O(\log t), \\
        \mathbf{a}_I=&O(1 / t^2),
    \end{aligned}
\label{vRa}
\end{equation}
where $\boldsymbol{V}_I$ is constant. In \cite{Gralla:2024wzr} the authors evaluate the total mass moment (\ref{massmoment}) and angular momentum (\ref{angmom}) on a constant-$\tau$ surface by using \eqref{ebfields} to \eqref{vRa} and find that the integrand of both quantities vanishes. Therefore, the scoot effect at 1PM order vanishes when the conserved quantities are evaluated on hyperboloidal slices. We now investigate whether a similar phenomenon occurs in the case of dyon scattering. For simplicity, let us start with one electric charge and one magnetic charge. Particle 1 has only an electric charge, so its electric and magnetic fields are still given by (\ref{ebfields}). On the other hand, particle-2 has only a magnetic charge, and its electric and magnetic fields are given by
\begin{equation}
    \begin{aligned}
        \mathbf{B}_{C, I}=&\left.\frac{g_I}{R_I^2} \frac{\hat{\mathbf{R}}_I-\mathbf{v}_I}{\left(1-v_I^2\right)^{-1}\left(1-\hat{\mathbf{R}}_I \cdot \mathbf{v}_I\right)^3}\right|_{\mathrm{ret}}, \\
       \mathbf{B}_{R, I}=&\left.\frac{g_I}{R_I} \frac{\hat{\mathbf{R}}_I \times\left((\hat{\mathbf{R}}_I-\mathbf{v}_I) \times \mathbf{a}_I\right)}{\left(1-\hat{\mathbf{R}}_I \cdot \mathbf{v}_I\right)^3}\right|_{\mathrm{ret}}, \\
       \mathbf{E}_{ I}=&-\left.\hat{\mathbf{R}}_I\right|_{\mathrm{ret}} \times \mathbf{B}_{I}.
    \end{aligned}
\label{ebfieldsformagcharge}
\end{equation}
Now we use (\ref{ebfields}) for particle-1, (\ref{ebfieldsformagcharge}) for particle-2 and construct the components of the electromagnetic energy-momentum tensor by using (\ref{emtensorcomps}). Following \cite{Gralla:2024wzr}, we only consider Coulomb fields, because they fall-off as $\tau^{-2}$ while radiation fields fall-off as $\tau^{-3}$ and looking at the forms of (\ref{massmoment}),(\ref{angmom}), only the product of Coulomb terms can survive the large-$\tau$ limit. Replacing (\ref{Rvector}) and (\ref{vRa}) into (\ref{ebfields}), electric and magnetic fields of particle-1 can be written as
\begin{equation}
    \boldsymbol{E}_1=A_1(\boldsymbol{r}-\boldsymbol{V}_1t) \quad \text{and} \quad  \boldsymbol{B}_1=-A_1(\boldsymbol{r} \times \boldsymbol{V}_1),
\end{equation}
where
\begin{equation}
    A_1=\frac{e_1}{R^3_1}\frac{(1-\boldsymbol{V}^2_1)}{(1-\hat{R}_1\boldsymbol{V}_1)^3}.
\end{equation}
Similarly, replacing (\ref{Rvector}) and (\ref{vRa}) into (\ref{ebfieldsformagcharge}), electric and magnetic fields of particle-2 can be written as
\begin{equation}
    \boldsymbol{B}_2=A_2(\boldsymbol{r}-\boldsymbol{V}_2t) \quad \text{and} \quad  \boldsymbol{E}_2=A_2(\boldsymbol{r} \times \boldsymbol{V}_2),
\end{equation}
where
\begin{equation}
    A_2=\frac{g_2}{R^3_2}\frac{(1-\boldsymbol{V}^2_2)}{(1-\hat{R}_2\boldsymbol{V}_2)^3}.
\end{equation}
Let us choose a frame where the particles are asymptotically collinear and their velocities are along the $z$-axis, namely,
\begin{equation}
    \boldsymbol{V}_1=V_1 \hat{\mathbf{z}}, \quad \boldsymbol{V}_2=-V_2\hat{\mathbf{z}}.
\end{equation}
With this choice, electric and magnetic fields for the particles are 
\begin{equation}
    \begin{aligned}
        \boldsymbol{E}_1=&A_1(x,y,z-V_1t), \quad \boldsymbol{B}_1=A_1(-V_1y,V_1x,0),\\
        \boldsymbol{E}_2=&A_2(-V_2y,V_2x,0), \quad \boldsymbol{B}_2=A_2(x,y,z+V_2t)
    \end{aligned}
\label{lastebfields}    
\end{equation}

Inserting these expressions (\ref{lastebfields}) into (\ref{emtensorcomps}), one can construct the electromagnetic energy-momentum tensor. Using this energy-momentum tensor in (\ref{massmoment}), one can easily show that the integrand vanishes and there is no contribution to the early/late time mass moment for the hyperboloidal slice boundary in monopole-electric charge scattering. This is the same result for the case of the scattering of two electric charges. Now, we continue with the calculation of angular momentum components using (\ref{angmom}). 

The integrand  in the calculation of angular momentum takes the form
\[
    I^{ij} = (x^iT^{j\mu}-x^jT^{i\mu})x_\mu .
\]

Taking into account the contributions of the first and second particles, the energy-momentum tensor, and hence $I^{ij}$,  naturally splits into three parts as
\begin{equation}\label{eq:integrand_split}
    I^{ij} = I^{ij}_{11}+I^{ij}_{22}+I^{ij}_{12}.
\end{equation}

In \eqref{eq:integrand_split}, the self-interaction terms $I^{ij}_{11}$ and $I^{ij}_{22}$ contain field contributions from only the first and second particles, respectively, whereas the cross-term $I^{ij}_{12}$ contains field contributions from both of the particles. \\

With the definitions given in \eqref{emtensorcomps}, $I^{ij}_{11}$ and $I^{ij}_{22}$ can be shown to vanish for all $i,j\in \{1,2,3\}$, while the cross-term contributions are given by 
\begin{widetext}
\begin{equation}
    I^{ij}_{12} = A_1A_2(-t^2+x^2+y^2+z^2)(V_1+V_2)
    \begin{pmatrix}
        0 & x^2+y^2 & yz \\
        -x^2-y^2 & 0 & -xz \\
        -yz & xz & 0
    \end{pmatrix} .
\end{equation}
\end{widetext}

The cross-term contribution to the late time angular momentum in electric charge-monopole scattering can be calculated by making use of \eqref{angmom}:
\begin{equation}
    M^{ij} = \int d\rho \theta d\phi \sinh^2\rho \sin\theta \; I^{ij}_{12}(\rho,\theta,\phi),
\end{equation}
where $I^{ij}_{12}(\rho,\theta,\phi)$ is evaluated using the coordinates given in \eqref{puzzle_coords}. Due to axial symmetry, the integrals of $I^{13}_{12}\sim \sin\phi$ and $I^{23}_{12}\sim \cos\phi$ over the $\phi$ coordinate vanish. The only non-vanishing piece is the $M^{12}=-M^{21}$ component, which unfortunately we were not able to compute. However, we expect a non-zero contribution to the change in the angular momentum in the z-direction,
\[
    \Delta L^z := M^{12}\bigg|_{\tau\to\infty} - M^{12}\bigg|_{\tau\to-\infty},
\] 
as we show later in \eqref{delta_Jz}.

\subsection{Particle Contribution}\label{sec4b}
In this section, we calculate the mechanical changes in the conserved quantities, when they are evaluated at the early/late proper
times (corresponding to the hyperboloidal slice boundary). To do this, we follow the analysis in \cite{Saketh:2021sri}, and extend this analysis for the scattering of an electric charge and a magnetic charge. \\

Consider the motion of two charged particles in classical relativistic electromagnetism in flat spacetime. The first particle have electric charge $e_1$ and mass $m_1$, while the second particle has magnetic charge $g_2$ and mass $m_2$. The particles have proper-time–parametrized worldlines given by $x^\mu_1 (\tau_1)$ and $x^\mu_2(\tau_2)$. The equation of motion for the worldline of the first particle is given by Lorentz equation
\begin{equation}
    m_1 \ddot{x}^\mu_1 = e_1 F_2^{\mu \nu}\dot{x}^\mu_1,
\label{eom1}
\end{equation}
where $F_2^{\mu \nu}$ is the field strength due to the field of the second particle, evaluated at the position of the first particle. The second particle has only magnetic charge in our construction, so the equation of motion for its worldline takes the form
\begin{equation}
    m_2 \ddot{x}^\mu_2 = g_2 \tilde{F}_1^{\mu \nu}\dot{x}^\mu_2
\label{eom2}
\end{equation}
where $\tilde{F}_1^{\mu \nu}$ is the dual field strength tensor due to the field of the first particle, evaluated at the position of the second particle. \\

Now we construct perturbative solutions to these equations and determine the particles' worldlines. At zeroth order, the worldlines can be taken as
\begin{equation}
    x_1^{(0) \mu}=b^\mu+u_1^\mu \tau_1, \quad x_2^{(0) \mu}=u_2^\mu \tau_2,
\end{equation}
where $u_1^\mu$ and $u_2^\mu$ are zeroth-order 4-velocities, $u_1.u_2 = \gamma$, $b^\mu$ is the impact parameter vector with $b \cdot u_1=0=b \cdot u_2$. \\

To solve equations (\ref{eom1}) and (\ref{eom2}), we need to determine $F_1^{\mu \nu}$ and $F_2^{\mu \nu}$. For the scattering of two electric charges, these field strength tensors are given in \cite{Saketh:2021sri}
\begin{equation}
    \begin{aligned}
        F_1^{\mu \nu}=&\frac{2 e_1}{r_{12}^3}\left(-b^{[\mu} u_1^{\nu]}+\tau_2 u_2^{[\mu} u_1^{\nu]}\right), \\
        F_2^{\mu \nu}=&\frac{2 e_2}{r_{21}^3}\left(b^{[\mu} u_2^{\nu]}+\tau_1 u_1^{[\mu} u_2^{\nu]}\right),
    \end{aligned}
\label{fieldtensors1}    
\end{equation}
where,
\begin{equation}
\begin{aligned}
    r_{21}=&r_2^{(0)}=\sqrt{|b|^2+(\gamma v)^2 \tau_1^2}, \\
    r_{12}=&r_1^{(0)}=\sqrt{|b|^2+(\gamma v)^2 \tau_2^2},
\end{aligned}
\end{equation}
and $v$ is the relative velocity between the particles. In our construction, the second particle has no electric charge, but has nonzero magnetic charge, so its field strength tensor changes as
\begin{equation}
    F_2^{\mu \nu} \rightarrow -\frac{ g_2}{r_{21}^3}\epsilon^{\mu \nu}{ }_{\sigma \rho}\left(b^{[\sigma} u_2^{\rho]}+\tau_1 u_1^{[\sigma} u_2^{\rho]}\right).
\end{equation}
One can also find the dual field strength tensor of the first particle from (\ref{fieldtensors1})
\begin{equation}
    \tilde{F}_1^{\mu \nu} = \frac{ e_1}{r_{12}^3}\epsilon^{\mu \nu}{ }_{\sigma \rho}\left(-b^{[\sigma} u_1^{\rho]}+\tau_2 u_2^{[\sigma} u_1^{\rho]}\right).
\end{equation}
So, the equations of the motions for the worldines of particles become
\begin{equation}
    \begin{aligned}
& m_1 a_1^{\mu}=-\frac{e_1 g_2}{r_{12}^3} u_{1 \nu} \epsilon^{\mu \nu}{ }_{\sigma \rho}\left(b^{[\sigma} u_2^{\rho]}+\tau_1 u_1^{[\sigma} u_2^{\rho]}\right), \\
& m_2 a_2 ^{\mu}=+\frac{e_1 g_2}{r_{21}^3} u_{2 \nu} \epsilon^{\mu \nu}{}_{\sigma \rho}\left(-b^{[\sigma} u_1^{\rho]}+\tau_2 u_2^{[\sigma} u_1^{\rho]}\right).
\end{aligned}
\label{a1vea2}
\end{equation}
Now, consider the initial rest frame of the second particle, 
\begin{equation}
    b^\mu=(0,b, 0,0), \quad u_1^\mu=\gamma(1,0,0, v), \quad u_2^\mu=(1,0,0,0).
\end{equation}
The acceleration four-vectors are found as
\begin{equation}
    \begin{aligned}
a^\mu_{1} & =\left(0,0, \frac{e_1 g_2 \gamma v b}{m_1\left(b^2+\gamma^2 v^2 \tau_1^2\right)^{3 / 2}}, 0\right), \\
a^\mu_{2} & =\left(0,0,-\frac{e_1 g_2 \gamma v b}{m_2\left(b^2+\gamma^2 v^2 \tau_2^2\right)^{3 / 2}}, 0\right)
\end{aligned}
\label{frame}
\end{equation}
Integrating once, one gets
\begin{equation}
    \begin{aligned}
        v^{(1) \mu}_{1}=&\left(0,0, \frac{e_1 g_2}{m_1 b}\left[1+\frac{\gamma v \tau_1}{\sqrt{b^2+\gamma^2 v^2 \tau_1^2}}\right], 0\right), \\
        v^{(2)\mu}_{2}= &\left(0,0,-\frac{e_1 g_2}{m_2 b}\left[1+\frac{\gamma v \tau_2}{\sqrt{b^2+\gamma^2 v^2 \tau_2^2}}\right], 0\right).
    \end{aligned}
    \label{velocities}
\end{equation}
Integrating once more, the first order corrections to the worldlines of the particles become
\begin{equation}
    \begin{aligned}
        x^{(1) \mu}_{1}=&\left(0,0, \frac{e_1 g_2}{m_1 b}\left[\tau_1+\frac{\sqrt{b^2+\gamma^2 v^2 \tau_1^2}}{\gamma v}\right], 0\right), \\
        x^{(1)\mu}_{2}=&\left(0,0,-\frac{e_1 g_2}{m_2 b}\left[\tau_2+\frac{\sqrt{b^2+\gamma^2 v^2 \tau_2^2}}{\gamma v}\right], 0\right).
    \end{aligned}
\label{trajectories}
\end{equation}
Now, using these trajectories we calculate the change in the angular momentum. Following \cite{Saketh:2021sri}, the change in angular momentum can be found by using
\begin{equation}
    \begin{gathered}
\Delta L^\mu=L_{\mathrm{f}}^\mu-L_{\mathrm{i}}^\mu, \\
L_{\mathrm{f} / \mathrm{i}}^\mu=\lim _{\tau_1, \tau_2 \rightarrow \pm \infty}-\epsilon^\mu{}_{\nu \rho \sigma} \frac{p_1^\nu p_2^\sigma}{E}\left(x_1-x_2\right)^\sigma, \\
p_a^\mu=m_a \dot{x}_a^\mu .
\end{gathered}
\label{angularmomentum}
\end{equation}
Inserting (\ref{fieldconserved}), (\ref{velocities}) and (\ref{trajectories}) into (\ref{angularmomentum}), we obtain
\begin{equation}
    \Delta L^x = \Delta L^y = 0, \quad \text{and} \quad \Delta L^z = \frac{2 e_1 g_2}{E}(\gamma m_1 + m_2).
\end{equation}
In the initial rest frame of the second particle (\ref{frame}), the total energy of the particles is $E=\gamma m_1 + m_2$, so we obtain,
\begin{equation}
    \Delta L^z = {2 e_1 g_2},\label{delta_Jz}
\end{equation}
which is the same result with (\ref{mechchange}). So we have shown that in the electric charge-magnetic charge scattering, the change in the total angular momentum of the particles is still there even if we formulate the conservation laws on hyperboloidal slice boundary.
\section{Discussion and Conclusion}
In the first part of this work, we have shown that the electromagnetic scoot—a permanent exchange of relativistic mass moment between particles and fields—extends naturally to the scattering of dyons. In \cite{emscoot}, the authors examine the relativistic scattering of two electric charges and show that there is a change in the total mass moment of the particles, which is balanced by an opposite change in the mass moment stored in the electromagnetic field. Their results show that this exchange is proportional to the product of charges $e_1$ and $e_2$. Our work is concentrated on the relativistic scattering of two dyons and we also observe this scoot effect for the mass moment. The net exchange we calculated depends both the electric charges and magnetic charges, i.e the proportionality factor we found is $e_1e_2+g_2g_2$. Besides the net exchange in relativistic mass moment, we also observe an exchange in angular momentum in the case of dyons. This is an expected result since Zwanziger already considered the scattering of dyons in \cite{zwanziger}, and found a residual angular momentum in the electromagnetic field of the in or out scattering states. While the calculations of Zwanziger are based on field theoretical methods, we consider a classical scattering process of two dyons and use classical equations of motion to derive the same result. Moreover, we extended the analysis of conserved quantities to the particles and explicitly showed that change in the angular momentum stored in electromagnetic field is balanced by an equal but opposite change in the total angular momentum of particles. Hence, we showed that the electromagnetic scoot phenomenon extends to the scattering process of dyons and provided an alternative route to derive the results of \cite{zwanziger}.\\

The second part of this work focuses on the scattering of an electric charge and a magnetic charge, and corresponding conservation laws formulated at hyperboloidal slice boundary. The motivation of this analysis comes from the results of \cite{Gralla:2024wzr}, where the authors consider the scattering of two electric charges and analyze the corresponding conserved quantities on hyperboloidal slice boundary. Their analysis shows that the electromagnetic scoot effect disappears at 1PM order for this boundary. We did the same analysis for the scattering of one electric charge with one magnetic charge to see if the electromagnetic scoot effect on the mass moment and angular momentum for dyons also disappears for such a boundary. For simplicity, we used one electric charge and one magnetic charge in our calculations, but the results can be easily generalized to the scattering of two dyons. In Sect.\ref{sec4a}, it has been showed that the field contribution to the early/late time mass moment vanishes for the hyperboloidal slice boundary in monopole-electric charge scattering. We also tried to find the field contribution to the early/late time angular momentum but could not find the result because of the complexity of the integral. So, in Sect.\ref{sec4b}, we analyzed the particle contribution to the early/late time angular momentum on hyperboloidal slice boundary for the electric-magnetic charge scattering process. The result of this section showed that there is a net change in total angular momentum given by 
\begin{equation}
    \Delta L^z = {2 e_1 g_2},
\end{equation}
which should be balanced with opposite contribution from the fields. So, this result shows that the dyonic contribution to the electromagnetic scoot effect for the angular momentum does not disappear at 1PM order, even if one
formulates the conservation laws on the hyperboloidal slice boundary. \\

The result of this work suggests that the dyonic contribution to the electromagnetic scoot effect is independent of the choice of boundary slicing. In this respect, it differs from the Coulombic contribution, which is sensitive to whether the conservation laws are evaluated on constant-time or hyperboloidal slices. Since these contributions are expected to be associated with additional quantum numbers in asymptotic multiparticle states, our findings may have implications for the representation theory of such states. We leave a detailed investigation of this connection to future work.
\begin{acknowledgments}

\end{acknowledgments}

We would like to express our deepest gratitude to Özgür Sarıoğlu for helpful discussions and a careful reading of this manuscript.

\bibliography{apssamp}

\end{document}